\documentclass{article}

\usepackage{graphics,latexsym}
\usepackage{graphicx}
\usepackage{amsmath, amssymb,amsthm}

\begin{document}

\title{Electromagnetic Fields  Simulating \\ a Rotating Sphere and its Exterior}
\author{Daniele Funaro}

\date{}

\author{ Daniele Funaro}
   
\maketitle

\centerline{\small Dipartimento di Scienze Chimiche e Geologiche} \centerline{Universit\`a di
Modena e Reggio Emilia}
\centerline{\small Via Campi 103, 41125
Modena (Italy)} \centerline{\small daniele.funaro@unimore.it}

\vspace{.5cm}

\begin{abstract}

Vector displacements expressed in spherical coordinates are proposed. 
They correspond to electromagnetic fields in vacuum that globally rotate
about an axis and display many circular patterns on the surface of a sphere. 
The fields basically satisfy the set of Maxwell's equations, but enjoy further properties that allow them to
be suitably interpreted as solutions
of a plasma model that combines electrodynamics with the Euler's equation for fluids.
Connection with magnetohydrodynamics can also be established.
The fields are extended with continuity outside the
 sphere in a very peculiar manner. In order to avoid peripheral velocities of arbitrary magnitude, as it may
happen for a rigid rotating body, they are organized to form successive encapsulated shells, with substructures recalling
successive ball-bearing assemblies. A recipe for the construction of these solutions is provided by
playing with the eigenfunctions of the vector Laplace operator. Some applications relative to astronomy are
finally discussed.
\end{abstract}

\smallskip
\noindent{Keywords:  exact solutions; electrodynamics; eigenfunctions; Bessel functions; associated Legendre polynomials.}
\par\smallskip

\noindent{AMS classification: 33C47,  	
35Q61,  	
78A25.  	
}

\par\medskip
\setcounter{equation}{0}

\section{Forewords}

The main practical achievement of this paper is the introduction of new exact solutions for a set of equations  modeling 
 electrodynamics. The result has by itself a general validity and can be applied in many circumstances emerging
in the context of electromagnetism (EM), magnetohydrodynamics (MHD), or geophysics. It may also represent a referring point to develop alternative numerical type simulations, especially 
in the context of spectral type approximation methods. 

The fields are described in spherical coordinates and solve the whole set of Maxwell's equations in vacuum.
They are obtained by separation of variables. As usual, this procedure leads to
trigonometric functions for the azimuthal angle and Bessel's functions for the radial component. As far as the altitude angle is concerned,
one obtains a family of special functions that can be put in connection with the so called {\sl Associated Legendre polynomials}. The proposed setting is not directly related to the family of {\sl vector spherical harmonics}.
The new fact here is that the displacement rigidly rotates about an axis at speeds comparable
to that of light, with an angular velocity depending on a parameter $\omega$.
The set up of the equations and the structure of the solutions is given in section 2. In the successive sections 3 and 4,  explicit
computations are carried out to check that the vector fields actually satisfy all the equations.

We also worked on the possibility to prolong the electromagnetic fields outside a given sphere. 
Since the solutions are defined in the whole tridimensional space, a natural extension already exists.
Nevertheless, such a straightforward expansion would bring to peripheral velocities of arbitrary magnitude,
which is unphysical. The question is however well-posed, since the dynamical fields present on the sphere
surface may be used as boundary constraints to analyze the external problem. Such a study is approached in \cite{landau}, section 89, from the relativistic viewpoint. Here, due to the lack of space, we will not touch on
questions pertaining to general relativity, although the argument is very appropriate.

\begin{center}
\begin{figure}[h!]\vspace{-.2cm}
\centerline{\includegraphics[width=13.5cm,height=4.5cm]{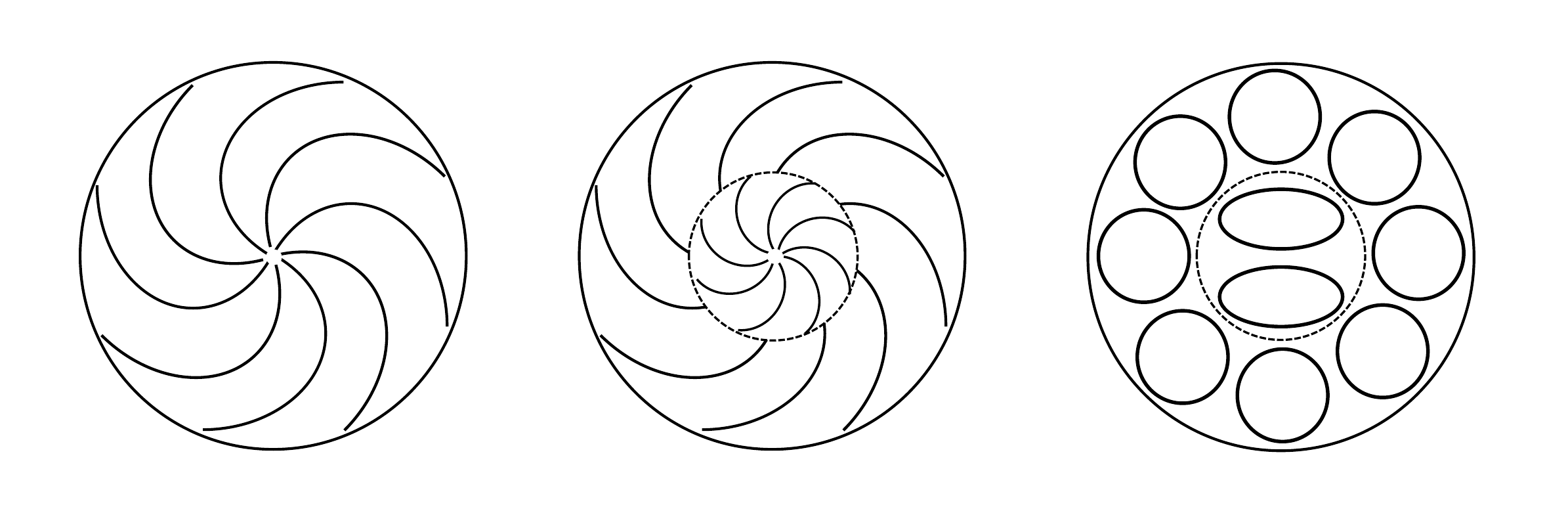}}
\vspace{-.3cm}
\begin{caption}{\small Due to the boundedness of the speed of light, the spiraling patterns of the left picture, tend to break up after a certain time
(central picture). In our approach (right), a fast rotating core induces the rotation of an external one with a lower angular velocity. Differently from
the previous case, this transfer can be done with continuity. }
\label{parker}
\end{caption}
\end{figure}
\end{center}
\vspace{-.2cm}

Because of the boundedness of the speed of light, as we move away from a central core, 
the rotatory information produces a spiral as in the first picture of Fig.  \ref{parker}.
Nevertheless, we guess that this behavior cannot be maintained for long times and sooner or later a breakdown occurs.
Such a phenomenon recalls that characterizing tornados, where a high-frequency whirl becomes isolated and drags the outside air to turn around at
lower frequency (secondary vortex). This is essentially due to the limits imposed by the speed of sound in air and by viscosity. 
We would like to reproduce a situation where the rotating sphere is surrounded by another EM configuration evolving at  lower angular velocity. 
The process may be repeated, so producing
a sequence of encapsulated environments. The most important achievement is that the connections can be done 
by avoiding discontinuities. This brings us to the third
picture of Fig. \ref{parker}, and to Fig. \ref{cuscinetto}. We postpone the discussion of the main idea to section 5,
by showing  how this can be quantitatively  implemented.

In section 6, we add stationary fields to the dynamical solutions examined so far, and we examine
the possible links with some model equations arising from the study of plasma. Finally, section 7
is devoted to some speculative considerations about the constitution of the Sun and the corresponding
solar system, that descend as a natural consequence of the analytic construction.

\section{Preliminary setting}

We start by introducing the classical Maxwell's equations in vacuum.
We denote by ${\bf E}=(E_1,E_2,E_3)$ the electric field and
by ${\bf B}=(B_1,B_2,B_3)$ the magnetic field. We first have
the Amp\`ere's law, with no current source term: 
\begin{equation}
\frac{\partial {\bf E}}{\partial t}~=~ c^2 {\rm curl} {\bf B}
\label{fem1}
\end{equation}
where $c$ is the speed of light.
Successively, we have the  Faraday's law:
\begin{equation}\label{fbm1}
\frac{\partial {\bf B}}{\partial t}~=~ -{\rm curl} {\bf E}
\end{equation}
Finally, we close the set with the following conditions on the divergence:
\begin{equation}\label{fde1}
{\rm div}{\bf E} ~=~0
\end{equation}
\begin{equation}\label{fdb1}
{\rm div}{\bf B} ~=~0
\end{equation}

It is well known that, by suitably combining the equations (\ref{fem1}), (\ref{fbm1}), (\ref{fde1}), 
(\ref{fdb1}), it is not difficult to arrive at the vector wave equations:
\begin{equation}\label{waveeb}
\frac{\partial^2 {\bf E}}{\partial t^2}~=~ c^2 ~\Delta {\bf E}
\qquad \qquad \frac{\partial^2 {\bf B}}{\partial t^2}~=~ c^2 ~\Delta {\bf B}
\end{equation}


It is standard to introduce the electromagnetic
\index{electromagnetic potentials} potentials ${\bf
A}=(A_1,A_2,A_3)$ and $\Phi$, such that:
\begin{equation}\label{potenz}
{\bf B}~=~\frac{1}{c}~{\rm curl}{\bf A}~~~~~~~~~{\bf E}~=~-\frac{1}{c}
\frac{\partial {\bf A}}{\partial t}~-~\nabla \Phi
\end{equation}
By assuming this, equations (\ref{fbm1}) and (\ref{fdb1}) are
automatically satisfied. The potentials are not unique, but are
usually related through some gauge condition. For convenience, the Lorenz's gauge will be assumed:
\begin{equation}\label{loreg}
{\rm div}{\bf A}~+~\frac{1}{c}\frac{\partial\Phi}{\partial t}~=~0
\end{equation}


We work in spherical coordinates $(r,\theta, \phi)$. Hereafter $H=H(r)$ will denote a function of the variable $r$,
whereas $S_2, S_3$ will be functions of the variable $x=\cos\theta$.
We also set $\zeta=c\omega t-m\phi $, where $\omega >0$ is a parameter. This setting allows us to simulate the
motion of an EM wave rotating
around the vertical axis (orthogonal to the equatorial plane $\theta =\pi/2$).
We then look for magnetic fields of the form:
\begin{equation}\label{campoB}
{\bf B}_D= (B_1, B_2, B_3)=\frac{1}{c}\Big( 0, \ H(r)S_2(\cos\theta )\cos\zeta, \ H(r)S_3(\cos\theta )\sin\zeta\Big)
\end{equation}
Thus, we have: $B_1=0$. The expression in  (\ref{campoB}) is required to satisfy the condition  ${\rm div}{\bf B}_D=0$
as well as the wave equation in (\ref{waveeb}). This is possible for special choices of the functions $H$, $S_2$, $S_3$.
The results of this tedious analysis are reported in the next sections. The subscript $D$ stands for {\sl Dynamical}, to
distinguish the present field from the {\sl Stationary} one ${\bf B}_S$ that will be introduced later on.


Regarding the electric field, it is enough to take the {\sl curl} of ${\bf B}_D$ and integrate with respect to time 
(see (\ref{fem1})). This yields:
$$
{\bf E}_D=(E_1, E_2, E_3)=\frac {1} {\omega}\left(\frac{H}{r}\sqrt{1-x^2}\left[ S'_3-\frac{x}{1-x^2}S_3+\frac{m}{1-x^2}S_2\right]\cos\zeta, \right.
$$
\begin{equation}\label{campoeg}
\qquad \qquad\left. \Big( H^\prime +\frac{H}{r}\Big)S_3\cos\zeta ,
\ \Big( H^\prime +\frac{H}{r}\Big)S_2\sin\zeta\right)
\end{equation}
where $S_3$ is differentiated with respect to $x$, and $H$ with respect to $r$.
Another equivalent version of (\ref{campoeg}) is found in (\ref{campoegg}).
It is a rather boring exercise to check that $\rho_D={\rm div}{\bf E}_D=0$ and that the electric field satisfies the vector wave
equation. It is interesting to point out that in general ${\bf E}_D\cdot{\bf B}_D\not= 0$, providing an example of solutions
of  Maxwell's equations where electric and magnetic fields  are not orthogonal.  
Note that, at the radial points where $H=0$, we obtain that ${\bf B}_D=0$ and that ${\bf E}_D$ is tangential to the sphere.
The distribution of the electric field in correspondence of these points is displayed in Fig. \ref{sfera}. 
As far as we know, the type of rotating waves we
are examining here are not related to the family of {\sl Vector Spherical Harmonics}.  Similarities
can be found with the eigenvalues of the {\sl Laplace's tidal equation} (see \cite{longuet}). The literature
offers plenty of results regarding the eigenmodes of the scalar Laplace equation in spherical coordinates, but
very few for the vector version.  Alternative solutions, naturally embedded in toroid shaped regions,  are explicitly found in \cite{chinosi} and  \cite{funaro6}. 


We can define the electromagnetic potentials by setting $\Phi_D=0$ and recovering ${\bf A}_D$ by integrating ${\bf E}_D$
in time. In this way the second relation in (\ref{potenz}) is fulfilled. Since ${\rm div}{\bf A}_D=0$ we are in the Lorenz's gauge. One can verify that also the first relation in (\ref{potenz}) is satisfied.


We conclude this section by defining the velocity vector field:
\begin{equation}\label{vu}
{\bf V}=\Big( 0, \ 0, \ \frac{c\omega}{m} r\sin \theta\Big)
\end{equation}
which actually simulates a uniform rotation around the vertical axis. An important relation that will be used later on
is the following one (see section 2 for the proof):
\begin{equation}\label{evb}
{\bf E}_D+{\bf V}\times {\bf B}_D=-\nabla p_D \ \ {\rm with} \ \  p_D=-\frac{1}{m\omega}( rH^\prime +H)S_2~ \sin\theta \cos\zeta
\end{equation}
This allows us to introduce a new potential $p_D$. Another solution, with the same properties of the one
just examined is found in \cite{funarol}, p. 147. Let us observe that in the homogeneous Maxwell's equations
the role of ${\bf E}_D$ and ${\bf B}_D$ can be interchanged. This is not true anymore if we want to preserve
the additional property (\ref{evb}).

\section{Explicit computation}

We first check that ${\bf B}_D$, as defined in (\ref{campoB}), has zero divergence and satisfies the vector
wave equation (\ref{waveeb}). More exactly, we show that:
\begin{equation}\label{eigb}
\frac{\partial^2 {\bf B}_D}{\partial t^2}=- c^2 \omega^2~{\bf B}_D \qquad 
\qquad \Delta {\bf B}_D=-\omega^2 {\bf B}_D
\end{equation}
The symbol $\Delta$ denotes the vector Laplacian in spherical coordinates. We recall that $S_2$ and $S_3$
are functions of $x=\cos \theta$. Since $B_1=0$, we must have:
$$
{\rm div }{\bf B}_D=\frac{1}{r\sin\theta}\left[ \frac{\partial}{\partial\theta}(B_2 \sin\theta )
+\frac{\partial B_3}{\partial \phi}\right]$$
\begin{equation}\label{divbc}
=\frac{H}{cr\sin\theta}\left[ \frac{d}{d\theta}(S_2 \sin\theta )
-m S_3\right]\cos\zeta =0
\end{equation}
Next, the following relations hold:
\begin{equation}\label{deri}
\frac{dS_2}{d\theta}=-S'_2(x)\sqrt{1-x^2}\qquad \qquad
\frac{d^2S_2}{d\theta^2}=S''_2(x) (1-x^2) -xS'_2(x)
\end{equation}
where the prime denotes the derivative with respect to $x$.
Thus, the divergence is zero everywhere if and only if:
\begin{equation}\label{deric}
\frac{d}{d\theta}(S_2 \sin\theta )-m S_3=-(1-x^2)S_2'+xS_2-mS_3 ~=~0
\end{equation}
which is equivalent to:
\begin{equation}\label{deric2d}
mS_3=-(1-x^2)S'_2+xS_2=-\sqrt{1-x^2}\Big(S_2 \sqrt{1-x^2}\Big)'
\end{equation}
We can then differentiate the above expression obtaining:
\begin{equation}\label{deric2}
mS'_3=-(1-x^2)S''_2+3xS'_2+S_2,\quad
mS''_3=-(1-x^2)S'''_2+5xS''_2+4S'_2
\end{equation}


As far as the eigenvalue problem in (\ref{eigb}) (second equation) is concerned, we
observe that the first component of $\Delta {\bf B}_D$ is automatically zero by
virtue of (\ref{divbc}):
\begin{equation}\label{lap1}
(\Delta {\bf B}_D)_1~=~-\frac{2}{r^2\sin\theta}\left[ \frac{\partial}{\partial\theta}(B_2 \sin\theta )
+\frac{\partial B_3}{\partial \phi}\right]~=~ 0
\end{equation}
Regarding the two other components, we have:
$$
c(\Delta {\bf B}_D)_2~=~\left( H''+\frac{2H'}{r}\right)S_2\cos\zeta 
$$
\begin{equation}\label{lap2}
 ~+~ \frac{H}{r^2}\cos\zeta\left[
\frac{1}{\sin\theta}~\frac{d}{d\theta}{\hspace{-.1cm}}\left( \sin\theta~\frac{dS_2}{d\theta}\right)-\frac{m^2+1}{\sin^2\theta}S_2
+\frac{2m\cos\theta}{\sin^2\theta}S_3\right]
\end{equation}

$$
c(\Delta {\bf B}_D)_3~=~\left( H''+\frac{2H'}{r}\right)S_3\sin\zeta 
$$
\begin{equation}\label{lap3}
 ~+~ \frac{H}{r^2}\sin\zeta\left[
\frac{1}{\sin\theta}~\frac{d}{d\theta}{\hspace{-.1cm}}\left( \sin\theta~\frac{dS_3}{d\theta}\right)-\frac{m^2+1}{\sin^2\theta}S_3
+\frac{2m\cos\theta}{\sin^2\theta}S_2\right]
\end{equation}
where here the prime denotes the derivative with respect to $r$.


We now require $H$ to satisfy the eigenvalue problem:
\begin{equation}\label{besselg}
H^{\prime\prime}+\frac{2H^\prime}{r}-\ell (\ell +1)\frac{H}{r^2}=-\omega^2 H
\end{equation}
The above equation leads us to spherical Bessel's functions.
Therefore, we have the linear combination of these expressions:
\begin{equation}\label{besseljyg}
H(r)=\sqrt{\frac{\pi}{2\omega r}}~J_{\ell +1/2}(\omega r)\qquad \qquad
H(r)=\sqrt{\frac{\pi}{2\omega r}}~Y_{\ell +1/2}(\omega r)
\end{equation}
where $J_\alpha$ is the Bessel's function of the first kind and $Y_\alpha$ is
that of the second kind. Note that the last one is singular for $r=0$. Note also that, for the moment, $\ell$
does not necessitate to be an integer.


By imposing that  $\Delta {\bf B}_D=-\omega^2 {\bf B}_D$, using  (\ref{deric2d}) and translating as a function 
of the variable $x$ the term in brackets of (\ref{lap2}), we arrive at:
\begin{equation}\label{eqf1}
(1-x^2)S''_2-4xS'_2 -\frac{m^2-1}{1-x^2}S_2 +[\ell (\ell +1)-2]S_2=0
\end{equation}
which is the equation to be satisfied by $S_2$. A discussion of the eigenvalue problem (\ref{eqf1}) is given in section 4.
We can argue with  (\ref{lap3}) in a similar way, obtaining:
\begin{equation}\label{eqf2}
(1-x^2)S''_3-2xS'_3 -\frac{m^2+1}{1-x^2}S_3 +\ell (\ell +1)S_3 +\frac{2mx}{1-x^2}S_2=0
\end{equation}
The above equation relates $S_2$ and $S_3$. We must observe at this point that a relation
between $S_2$ and $S_3$ has been already set up through (\ref{deric2d}). We have to check that these
two are equivalent. To this end, we differentiate (\ref{eqf1}) with respect to $x$: 
\begin{equation}\label{eqdif}
(1-x^2)S'''_2-6xS''_2 -\frac{m^2-1}{1-x^2}S'_2 -\frac{2(m^2-1)x}{(1-x^2)^2}S_2+[\ell (\ell +1)-6]S'_2=0
\end{equation}
Using (\ref{deric2d}) we rewrite (\ref{eqf2}) in terms of the sole unknown $S_2$, which appears
with its third derivative $S'''_2$. This last is replaced by that recovered from (\ref{eqdif}), in order to get a second
order differential equation. After tedious simplifications one discovers that this last equation is nothing but 
(\ref{eqf1}) multiplied by $x$. Hence, by assuming that both (\ref{deric2d}) and (\ref{eqf1}) are true, we
actually verify (\ref{eqf2}), showing that $(\Delta {\bf B}_D)_3=-\omega^2 B_3$ is automatically satisfied 
when  ${\rm div }{\bf B}_D=0$ and  $(\Delta {\bf B}_D)_2=-\omega^2 B_2$ hold true.
We can finally claim that the wave equation for ${\bf B}_D$ is satisfied when  $H$ comes
from (\ref{besselg}), $S_2$ comes from $(\ref{eqf1})$, and $S_3$ is chosen as in (\ref{deric2d}). 


As a final exercise we check (\ref{evb}). We start by writing:
\begin{equation}\label{vb}
{\bf V}\times {\bf B}_D= -\frac{\omega}{m}\Big( Hr S_2 \sin\theta \cos\zeta , \ 0, \ 0\Big)
\end{equation}
We have to prove that:
\begin{equation}\label{gradp}
-\nabla p_D =-\left( \frac{\partial p_D}{\partial r}, \ \frac{1}{r}\frac{\partial p_D}{\partial \theta},
 \ \frac{1}{r\sin\theta}\frac{\partial p_D}{\partial \phi}\right)= \Big( ({\bf E}_D+{\bf V}\times {\bf B}_D )_1,
 E_2, E_3\Big)
\end{equation}
The verification is straightforward for the second and the third components. Concerning the first one, we
begin with noting that:
$$\ell (\ell+1) S_2 
 =-(1-x^2)S''_2+4xS'_2 +\frac{m^2-1}{1-x^2}S_2 +2S_2
$$
\begin{equation}\label{expl}
=m \left[ S'_3-\frac{x}{1-x^2}S_3+\frac{m}{1-x^2}S_2\right]
\end{equation}
where we used (\ref{eqf1}), the first equation in (\ref{deric2}), and (\ref{deric2d}).
The last term above is equal to the one in square brackets in (\ref{campoeg}).
We finally have:
$$
-\frac{\partial p_D}{\partial r} =\frac{1}{m\omega}(rH''+2H') S_2 \sin\theta \cos\zeta
=\frac{H}{m\omega}\left(\frac{1}{r}\ell (\ell+1)-\omega^2 r\right) S_2 \sin\theta \cos\zeta
$$
\begin{equation}\label{gradpv}
=\frac{H}{m\omega r}\ell (\ell+1) S_2 \sin\theta \cos\zeta +({\bf V}\times {\bf B}_D)_1 = ({\bf E}_D+{\bf V}\times {\bf B}_D )_1
\end{equation}
where we used (\ref{besselg}). This completes the proof of (\ref{gradp}).


By virtue of (\ref{expl}), we can rewrite (\ref{campoeg}) as follows:
\begin{equation}\label{campoegg}
{\bf E}_D\hspace{-.1cm}=\hspace{-.1cm}\frac {1}{\omega}\hspace{-.1cm}\left(\frac{H}{rm}\ell (\ell +1) S_2\sin\theta \cos\zeta, 
\Big(\hspace{-.1cm} H^\prime +\frac{H}{r}\Big)S_3\cos\zeta ,
\ \Big(\hspace{-.1cm} H^\prime +\frac{H}{r}\Big)S_2\sin\zeta\right)
\end{equation}


\section{Further insight}

Here, we would like to examine the properties of equation (\ref{eqf1}).
As a particular example, we start by studying the case $m=1$, where  $\zeta =c\omega t-\phi$.
In this situation the set of solutions is well characterized. Up to multiplicative constants, it consists of the 
set of Jacobi polynomials $P^{(1,1)}_\ell$,
where $\ell$ is integer and denotes the degree. For instance, by taking $\ell =1$, we get $S_2(x)=1$ and $S_3(x)=x$.
In this simple situation the vector potential takes the form:
\begin{equation}\label{potv}
{\bf A}_D=-\frac{1}{\omega^2}\left(\frac{2}{r}H\sin\theta\sin\zeta, \ \Big( H^\prime +\frac{H}{r}\Big)\cos\theta\sin\zeta ,
\ -\Big( H^\prime +\frac{H}{r}\Big)\cos\zeta\right)
\end{equation}
where $H$ satisfies:
\begin{equation}\label{bessel}
H^{\prime\prime}+\frac{2H^\prime}{r}-\frac{2H}{r^2}=-\omega^2 H
\end{equation}
having for solution a linear combination of the two following Bessel's functions:
\begin{equation}\label{besseljy}
\sqrt{\frac{\pi}{2\omega r}}~J_{3/2}(\omega r)=\frac{\sin \omega r}{(\omega r)^2}-\frac{\cos \omega r}{\omega r}, 
\quad \sqrt{\frac{\pi}{2\omega r}}~Y_{3/2}(\omega r)=-\frac{\cos \omega r}{(\omega r)^2}-\frac{\sin \omega r}{\omega r}
\end{equation}

Note that the vector representing the magnetic field ${\bf B}_D$ at the poles ($\theta =0$ or $\theta =\pi$) is tangent
to the sphere and rotates with angular velocity $c\omega$. This will not be true for $m>1$, where ${\bf B}_D$ is going
to be zero at the poles.

\begin{center}
\begin{figure}[h!]\vspace{-.2cm}
\centerline{\includegraphics[width=6.8cm,height=5.1cm]{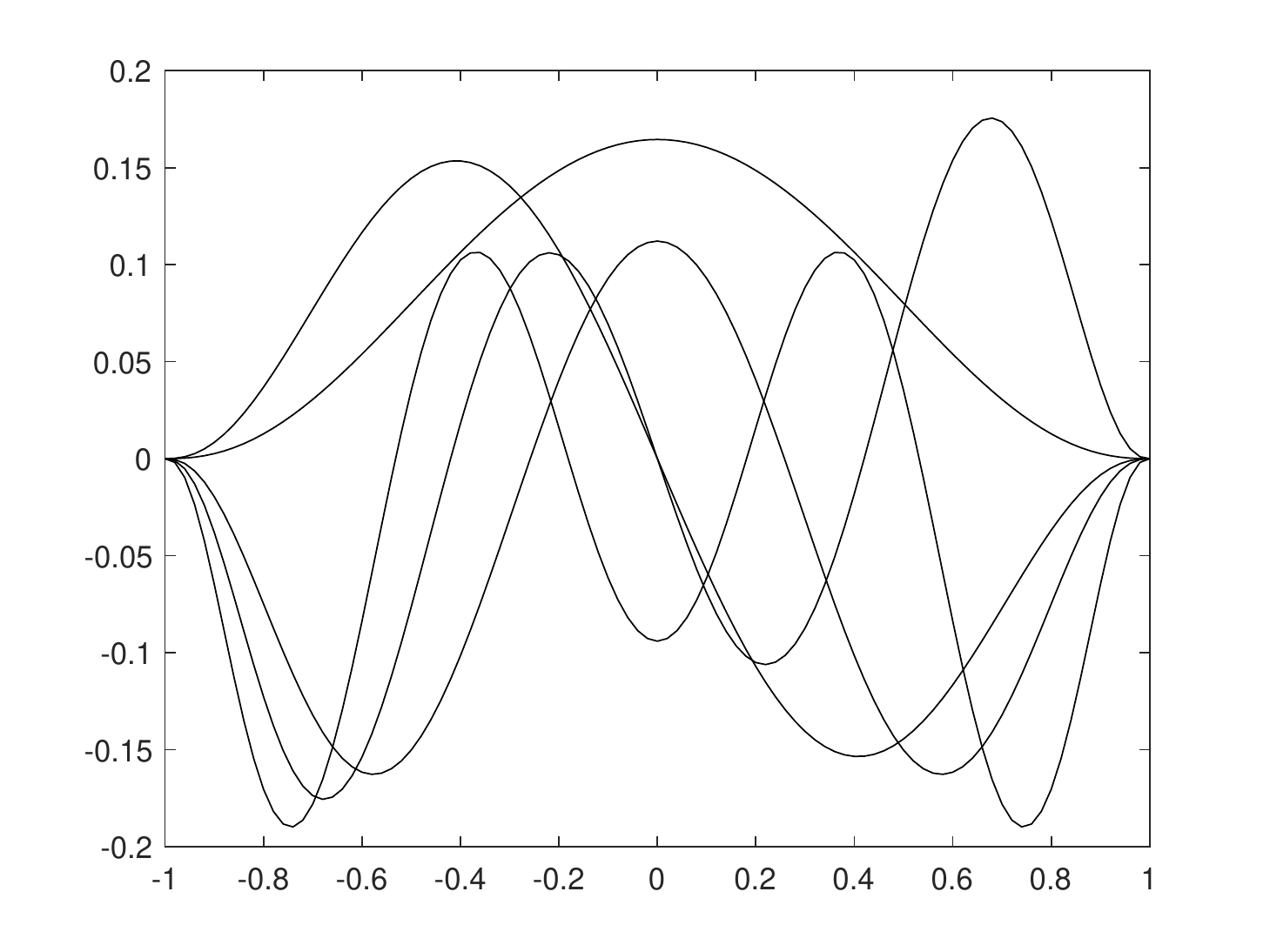}
\hspace{-.8cm}
\includegraphics[width=6.8cm,height=5.1cm]{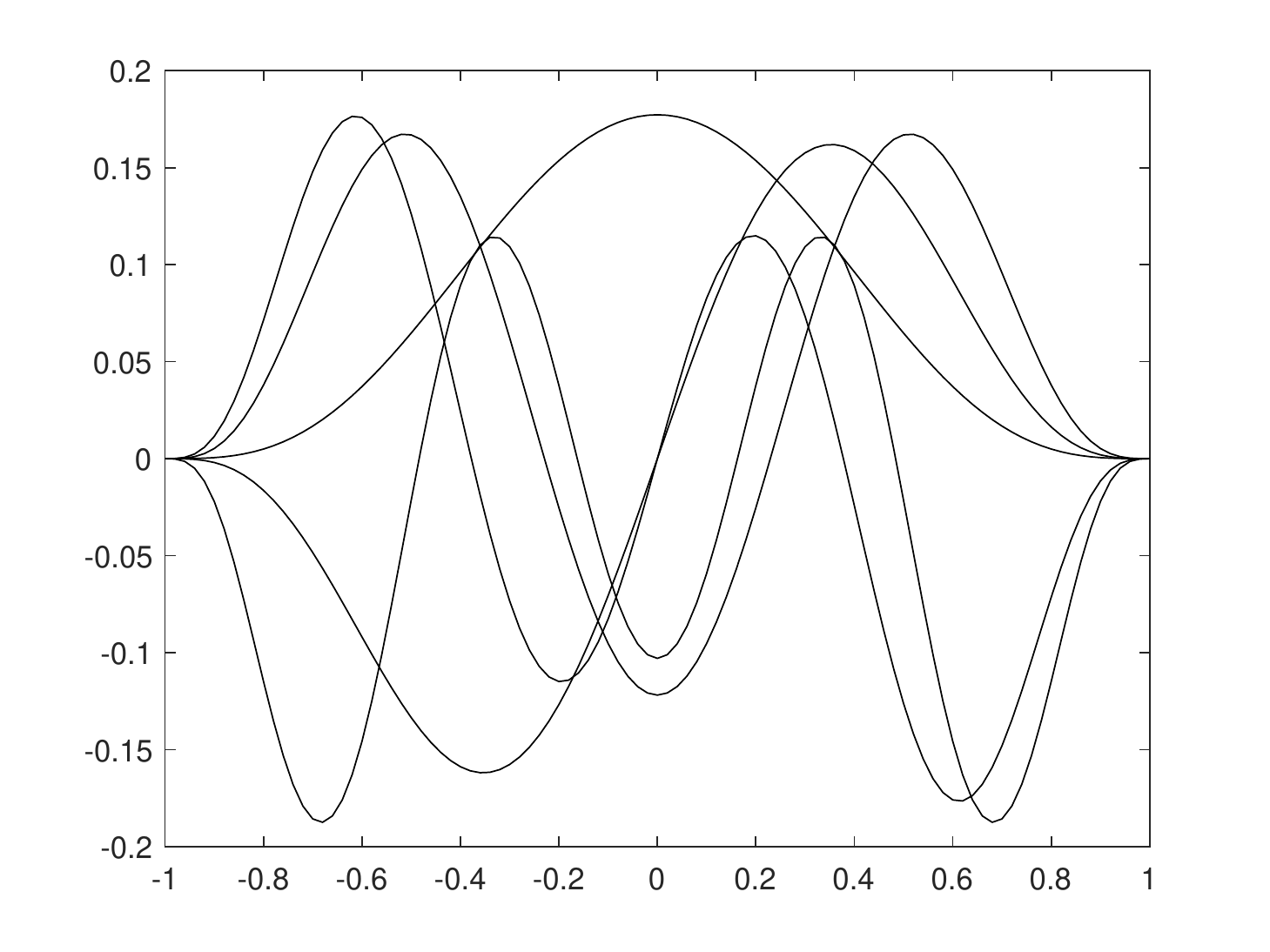}}
\vspace{-.3cm}
\begin{caption}{\small Eigensolutions of (\ref{eqf1}) for $m=6$, $\ell = 6, 7, 8, 9, 10$ (left)
and $m=8$, $\ell =8, 9, 10, 11, 12$ (right).}
\label{autofunzioniS}
\end{caption}
\end{figure}
\end{center}\vspace{-.4cm}

We did not find documentation for a general $m>1$, but we presume that the problem has been
already studied. The situation is very similar to that related to the so called {\sl Associated 
Legendre polynomials} (see e.g. \cite{abra}, p. 331), where the term $-m^2/(1-x^2)$ is added to the classical differential operator
$(1-x^2)(d^2/dx^2)-2x(d/dx)$. Here the operator is slightly different, i.e.: $(1-x^2)(d^2/dx^2)-4x(d/dx)$.
Connections with the so called Hough functions may be devised (see, e.g.: \cite{wang}).
The results of some numerical computations show that the eingenvalue problem:
\begin{equation}\label{eif1}
(1-x^2)S''-4xS' -\frac{m^2-1}{1-x^2}S =-\lambda S
\end{equation}
is actually solvable. The values of $\lambda$ exactly correspond to numbers of the form $[\ell (\ell +1)-2]$,
where $\ell$ is integer with $\ell \geq m$. For a fixed $m>1$ the eigenfunctions increase their
frequency as $\ell$ grows. Differently from the case $m=1$, they satisfy homogeneous boundary conditions at the points $\pm 1$ (the poles of the sphere). Some plots are given in Fig. \ref{autofunzioniS}.
The eigenfunctions are orthogonal with respect to the inner product of the space $L^2_w(-1,1)$, weighted by the function $w=1-x^2$.
A similar orthogonality relation holds for the derivatives. In particular, if $S$ and $T$ are solutions of (\ref{eqf1}) corresponding to different
values of $\ell$, we have:
\begin{equation}\label{ort}
\int_{-1}^1 S T (1-x^2)dx=0, \quad \int_{-1}^1 S'T' (1-x^2)^2dx +(m^2-1)\int_{-1}^1 S T dx =0
\end{equation}

\begin{center}
\begin{figure}[h!]\vspace{-.1cm}
\centerline{\includegraphics[width=8.4cm,height=8.cm]{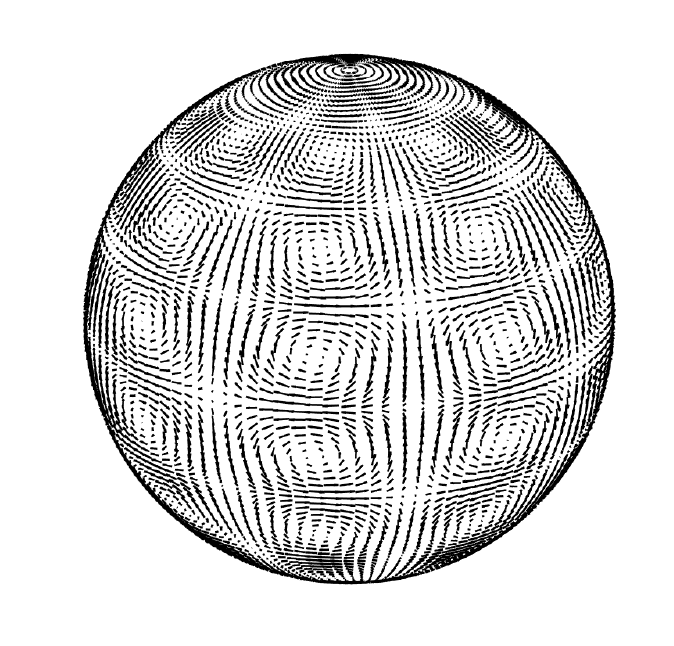}}
\vspace{-.2cm}
\begin{caption}{\small Electric field distribution on the surface of a sphere having the radius corresponding
to a zero of the function $H$. In this situation, the magnetic field is uniformly zero. The number of vortexes depends on the parameters.
In the present case we have: $m=4$ and $\ell =11$.}
\label{sfera}
\end{caption}
\end{figure}
\end{center}

The distribution of the electromagnetic fields is rather complicated. We can examine the case of a sphere having the radius
corresponding to a zero of $H$ (recall that $H$ is related to Bessel's functions, so that it displays infinite zeros).
By looking at (\ref{campoB}) and (\ref{campoeg}), we easily realize that ${\bf B}_D$ is identically zero on the surface of that sphere,
whereas ${\bf E}_D$ turns out to be tangential and organized to form several vortexes. The number of vortexes along the azimuthal direction
is ruled by the parameter $m$. The number of vortexes spanned by the altitude angle depends on $\ell$.
A typical configuration is displayed in Fig. \ref{sfera}. The displacement rotates about the vertical axis, as
prescribed by the velocity field ${\bf V}$.
Similar conclusions can be made for ${\bf B}_D$ when the radius of the sphere is not equal to a zero of $H$.

\vspace{.1cm}

\begin{center}
\begin{figure}[h!]\vspace{-.1cm}
\centering
\includegraphics[width=13.5cm,height=7.7cm]{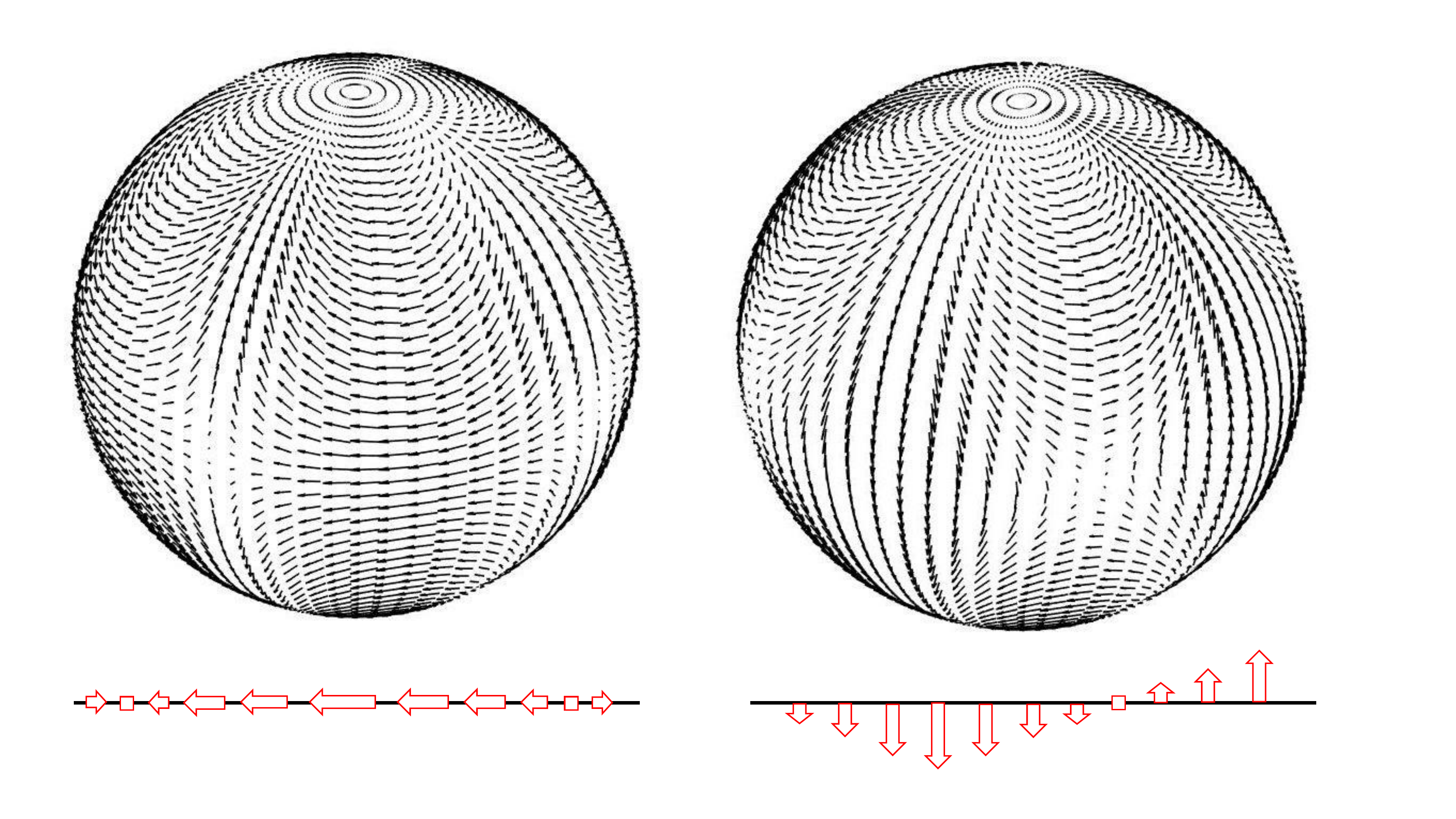}
\vspace{-1.cm}
\begin{caption}{\small Field distributions: electric (left) for radial points where $H=0$; magnetic (right) for radial points where $H\not =0$. Both pictures
refer to the situation  $m=\ell =2$. 
At the equator, the electric field oscillates horizontally, whereas the magnetic field is lined up with the meridians.}
\label{sfere}
\end{caption}
\end{figure}
\end{center}

An explicit solution for $m=\ell \geq 1$ is known. This is given by setting:
\begin{equation}\label{casomell}
S_2(\cos\theta )= (\sin \theta )^{m-1}\qquad \qquad S_3(\cos\theta )= \cos\theta (\sin \theta )^{m-1}
\end{equation}
With the help of (\ref{casomell}) we can better examine the distribution of the fields in the case $m=\ell =2$. For $H\not =0$, we get from  (\ref{campoB}):
\begin{equation}\label{mell2b}
{\bf B}_D=\frac{H(r)}{c}\Big( 0, \ \sin\theta \cos \zeta ,  \ \sin\theta \cos\theta \sin \zeta \Big)
\end{equation}
For $H=0$, we get instead from  (\ref{campoeg}):
\begin{equation}\label{mell2e}
{\bf E}_D=\frac{H'(r)}{\omega}\Big( 0, \ \sin\theta \cos\theta \cos \zeta ,  \ \sin\theta  \sin \zeta \Big)
\end{equation}
The corresponding electric and magnetic fields are respectively shown in Fig. \ref{sfere}.

\vspace{-.6cm}


\section{Extension outside the sphere}

We continue our exploration on the solutions (\ref{campoB}) and (\ref{campoeg}) in the special case where $m=\ell$.
We examine what happens outside the sphere.
The expressions in (\ref{casomell}) tell us that, for large $m$, the function $S_2$ assumes the 
value 1 for $\theta =\pi /2$ and goes fast to zero as approaching $\theta =0$ or $\theta =\pi$.
The fields are then concentrated in a flat annular region around the equator. As reported in Fig. \ref{sfere}, for $H=0$, the electric field is 
lined up with the equator and oscillates according to the rule $\cos\zeta =\cos (c\omega t- m\phi )$.
For $H\not =0$, the magnetic field follows the same behavior, but it is oriented as the meridians.
Based on (\ref{vu}), at a radius $r$, the intensity of the peripheral velocity of this equatorial wave is $V(r)=c\omega r/m$.


As $r$ reaches a zero of $H$, the magnetic field vanishes. We believe that in this circumstance there is
a change of regime. In fact, it is difficult to accept the idea that, by increasing $r$, the quantity $V(r)$ is allowed to assume any possible
value, as it would be for a rotating rigid body in classical mechanics. It is reasonable to guess instead that $V(r)$ does
not exceed too much the speed of light in vacuum. This is thinkable if we suppose that the process happens through some quantized steps.
Indeed, it is possible to build encapsulated shells. Inside each one of them we are solving a wave type equation. The angular
velocity decreases by passing from a shell to an external one. Moreover, such a passage can be made in continuous way.
We show how this extraordinary fact can be achieved.


We adapt to the present circumstances a situation already studied in \cite{funaro4}. The aim is to construct solutions defined
on a circular crown, in such a way that the velocity at the internal boundary is more or less the same (in magnitude)
than that at the external boundary. This construction relies on the possibility to find eigenfunctions of the Laplace
operator, corresponding to eigenvalues of multiplicity four, at least. 
Technically, the question is reduced to find suitable periodic solutions of the wave equations  (\ref{waveeb}). After separation of variables and
further simplifications ($m=\ell $), the problem can be studied for a scalar equation in two dimensions, but the general discussion in 3D involves the same ingredients.
The domain is an annular region between the radii $r_{\rm min}$ and  $r_{\rm max}$. Homogeneous Dirichlet boundary conditions are
assumed, though such a constraint is not strict. The solution must develop in such a way that, in proximity of the inner boundary, the shift is governed
by the rule $\cos (c\omega t-m_A\phi )$, where $m_A$ is an integer. At the outside boundary we should have instead $\cos (c\omega t-m_B\phi )$,
where $\vert m_B\vert >\vert m_A\vert $ is another integer. At these boundaries, the velocity of rotation is expressed by  (\ref{vu}). If the rotating body was rigid,
the external velocity would be larger than the internal one, directly depending on the ratio   $r_{\rm max}/r_{\rm min}$. Here we can play instead
with the values of the integers $m_A$ and $m_B$, in order to obtain that the intensity of the inner velocity $c\omega r_{\rm min}/m_A$
is comparable with the external one $c\omega r_{\rm max}/m_B$.

Such an analysis is not trivial and passes through the determination of the zeros of the Laplacian eigenfunctions in the domain. In fact, not
all the configurations are possible. The parameters to play with are:  $m_A$, $m_B$, $\omega$ and $r_{\rm max}/r_{\rm min}$. 
They have to be detected in order to have a basis of at least four orthogonal eigenfunctions corresponding to the same eigenvalue
(which, as a consequence, must have multiplicity equal to 4). Interesting dynamical patterns are then obtained from suitable linear combinations of these eigenfunctions.

\begin{center}
\begin{figure}[h!]\vspace{.2cm}
\centerline{\includegraphics[width=6.5cm,height=6.cm]{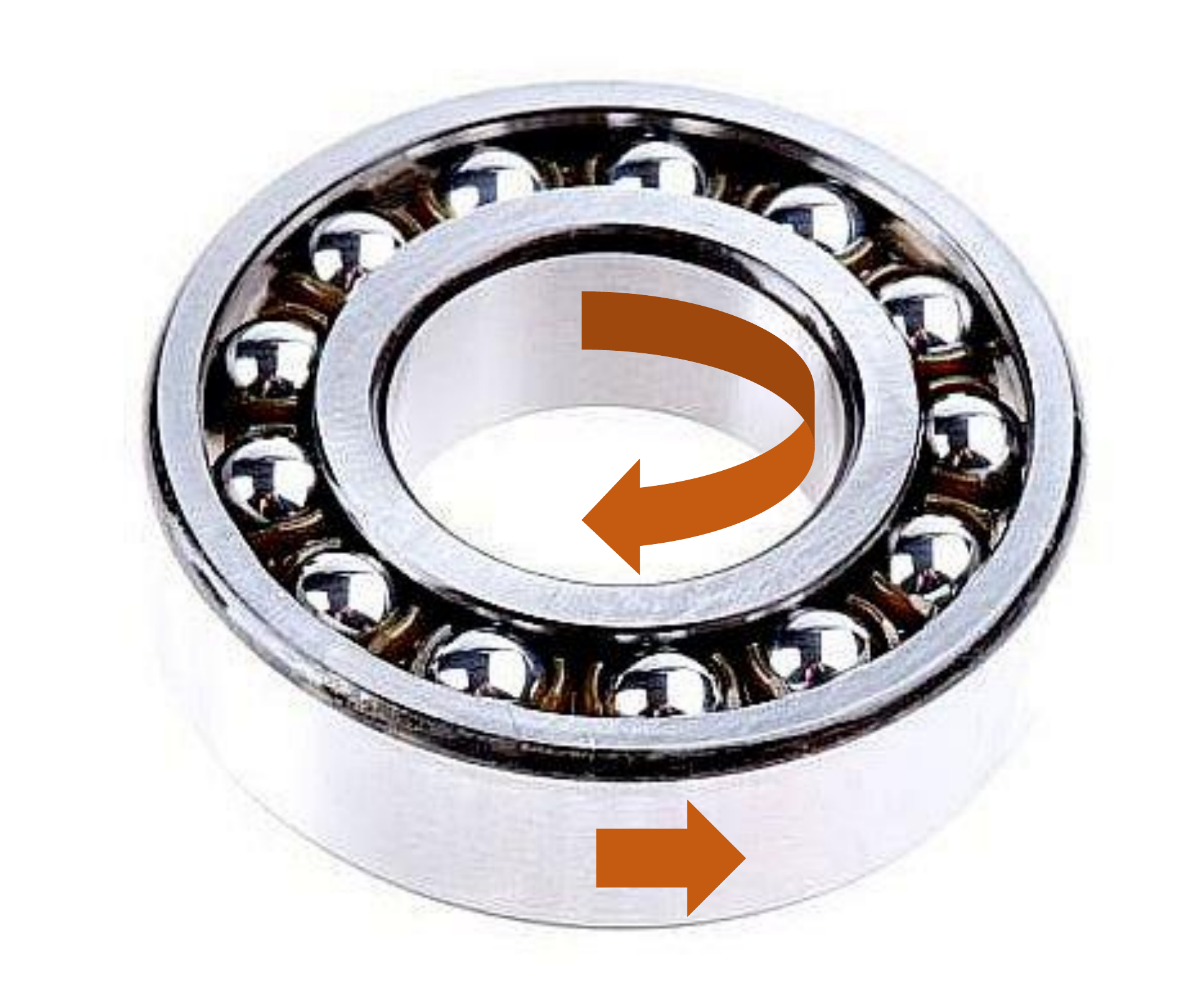}\hspace{.1cm}
\includegraphics[width=6.1cm,height=5.6cm]{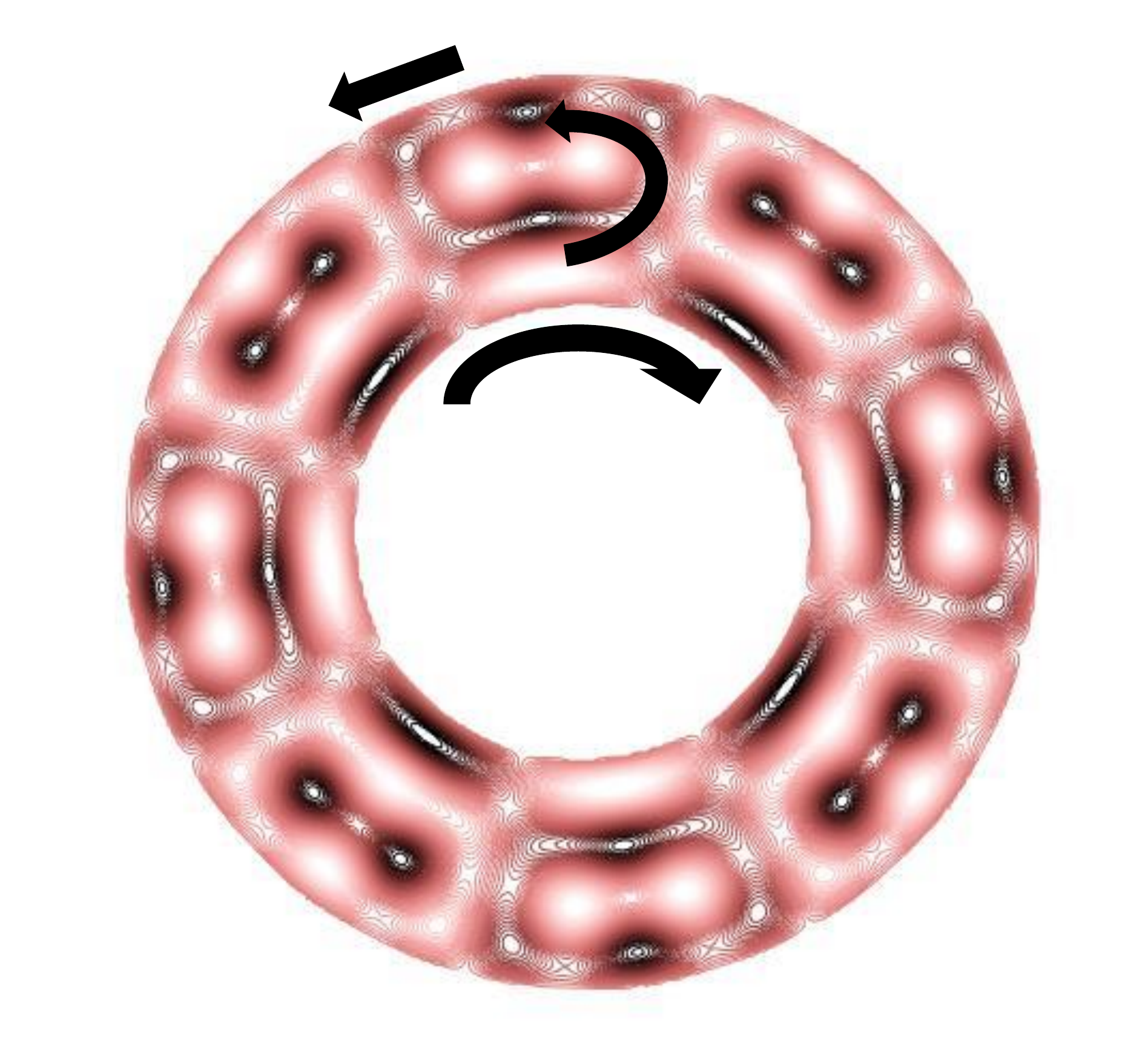}}
\vspace{-.2cm}
\begin{caption}{\small Ball bearing assembly. In the referring frame where the spheres are at rest, the internal and the external boundaries
counter rotate, displaying different angular velocities. A similar effect can be achieved by solving the wave equation in a suitable annular region.
In this circumstance, the size and the frequencies involved must be wisely calibrated.}
\label{cuscinetto}
\end{caption}
\end{figure}
\end{center}

The underlying idea of this construction is to recreate something similar to an interconnected set of gears
of different size: the small one turning fast, imparts a slow rotation to the big one. The case of an annular region is better described
by a ball bearing assembly (see Fig. \ref{cuscinetto}), where, in smooth way, the momentum of the internal support is transferred to the external one,
avoiding the inconveniences (and the paradoxes) related to the rotation of a rigid body. 

From the practical viewpoint, let  $r_{\rm min}$ denote the interior radius of the 
annulus.  By recalling  (\ref{besseljyg}), we define the following function:
\begin{equation}\label{effem}
F_m(r ) =\frac{1}{\sqrt{\omega r}}\Big( Y_{m+\frac12}(\omega r_{\rm min}) J_{m+\frac12}(\omega r)
-J_{m+\frac12}(\omega r_{\rm min}) Y_{m+\frac12}(\omega r)\Big)
\end{equation}
It is easy to check that: $F_m( r_{\rm min} )=0$.

\vspace{-.2cm}

\begin{center}
\begin{figure}[h!]\vspace{-.6cm}
\centerline{\includegraphics[trim = 5cm 10cm 4cm 10cm, clip, scale=.65]{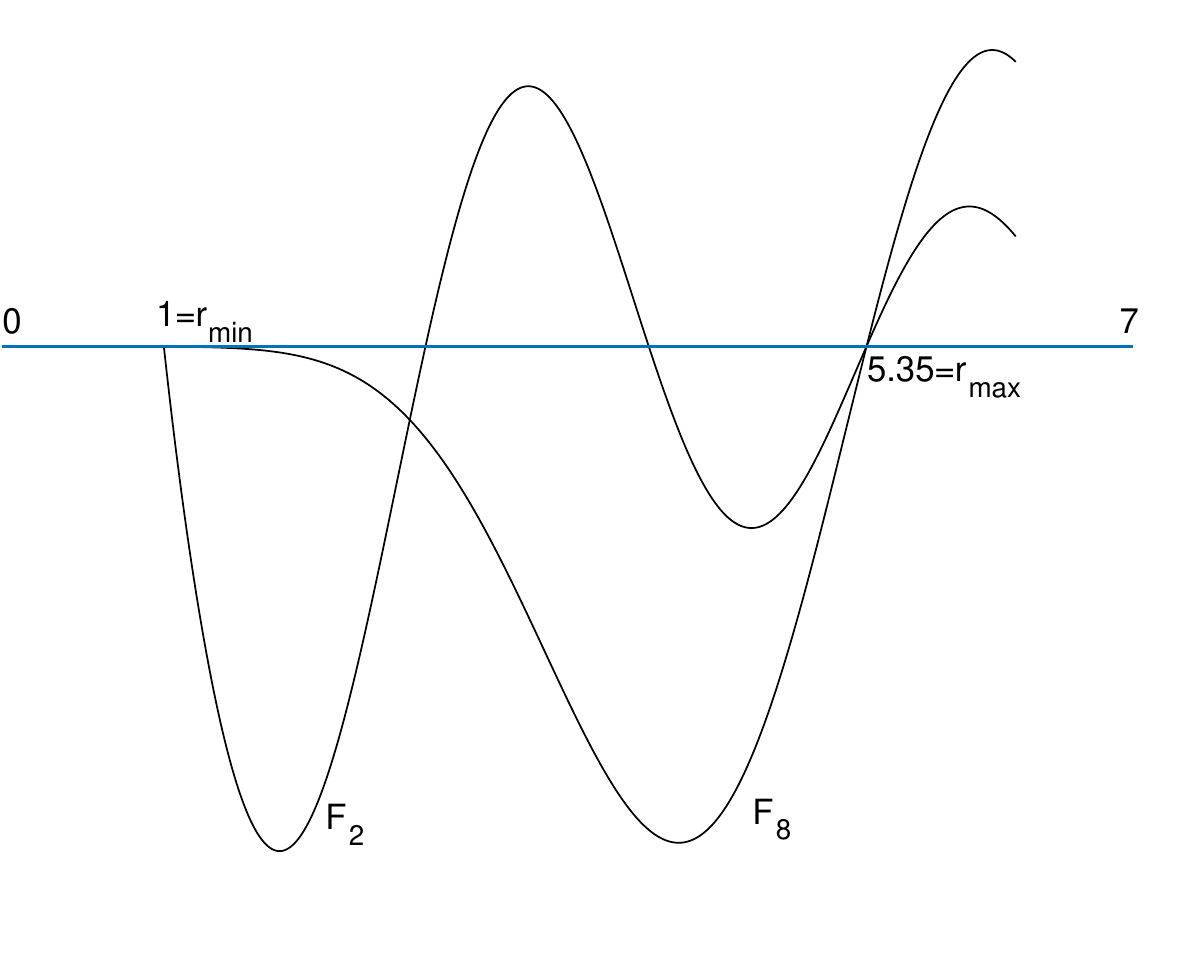}}
\vspace{-.3cm}
\begin{caption}{\small Both the two functions $F_2$ and $F_8$ vanish at $r_{\rm min}=1$ and  $r_{\rm max}\approx 5.35$.
The amplitude of the functions have been suitably rescaled to make more clear the graphical output.}
\label{besseldoppio}
\end{caption}
\end{figure}
\end{center}

We would like now to find two different integers  $m_A$ and  $m_B$, a value of the parameter $\omega$,  and a radius  $r_{\rm max}$ of the external 
circumference of the annulus. This has to be done in order to satisfy the conditions:
\begin{equation}\label{effemax}
F_{m_A}( r_{\rm max})=0, \qquad  \qquad F_{m_B}( r_{\rm max} )=0
\end{equation}
The explanation is as follows. We require homogeneous Dirichlet conditions on the boundaries of the annulus (internal and external), and we
want this to be simultaneously achieved for different frequencies $m_A$ and  $m_B$. Such a problem does not always admit solution. 
Possible allowed combinations (among infinite others) for $r_{\rm min}=1$ are: $m_A=2, m_B=5, \omega \approx 1.97,  r_{\rm max}\approx 4.75$;  $m_A=2, m_B=6, \omega \approx 3.72,  r_{\rm max}\approx 2.83$;
or $m_A=2, m_B=8, \omega\approx 2.39,  r_{\rm max}\approx 5.35 $. The last case is represented in Fig. \ref{besseldoppio}.

\vspace{-.3cm}

Reminding once again that we are examining the case $m=\ell$ (see also
(\ref{casomell})), the expression in (\ref{effem}) has a link with the eigenfunctions studied so far, where the
dependence from  the variable $\theta$ is neglected. We can actually define:
\begin{equation}\label{evolfg}
\Phi_m (r, \phi )=\alpha_m F_m(r) \cos (m\phi )\qquad\qquad
\Psi_m (r,\phi) =\beta_m F_m(r) \sin (m\phi )
\end{equation}
for arbitrary multiplicative constants $\alpha_m$ and $\beta_m$.
By restoring the part in the variable $\theta$, these are indeed two orthogonal eigenfunctions with eigenvalue $-\omega^2$. 
We get solutions of the wave equation by introducing combinations depending on time. For different values of 
$m_A$ and  $m_B$, we can write:
\begin{equation}\label{eigfg}
\Phi_{m_A}\sin (c\omega t) + \Phi_{m_B}\cos (c\omega t)
+\Psi_{m_A}\sin (c\omega (t +t_0)) + \Psi_{m_B}\cos (c\omega (t+t_0))
\end{equation}
where $t_0$ is a time shift. 

\begin{center}
\begin{figure}[p!]\vspace{.0cm}
\centerline{\includegraphics[trim = 6cm 10cm 5.5cm 10cm, clip, scale=0.55]{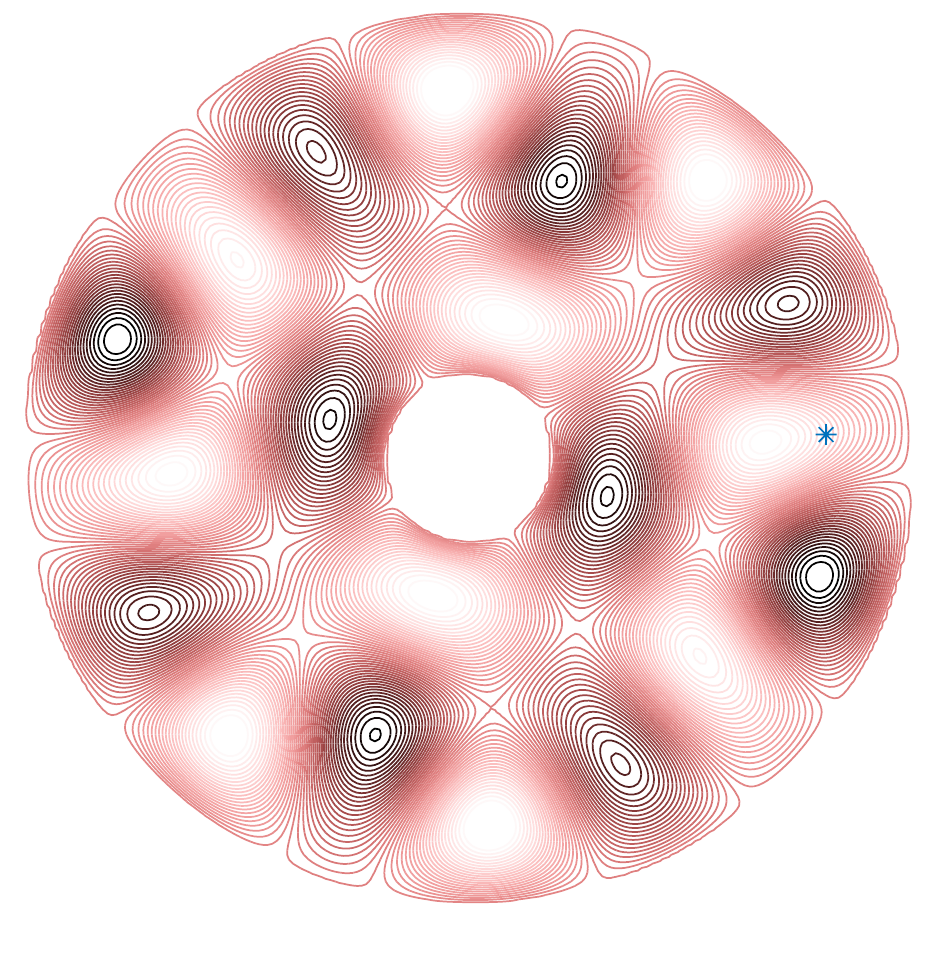}\hspace{.7cm}
\includegraphics[trim = 6cm 10cm 5.5cm 10cm, clip, scale=0.55]{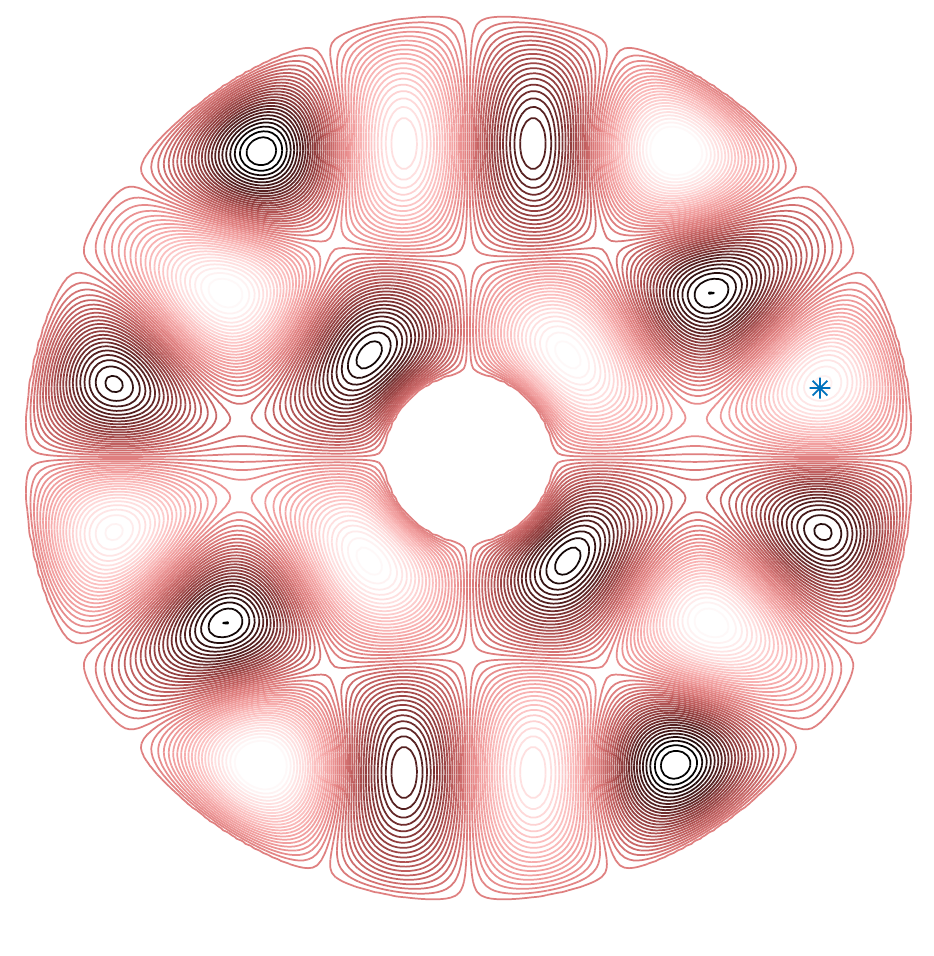}}
\centerline{\includegraphics[trim = 6cm 10cm 5.5cm 10cm, clip, scale=0.55]{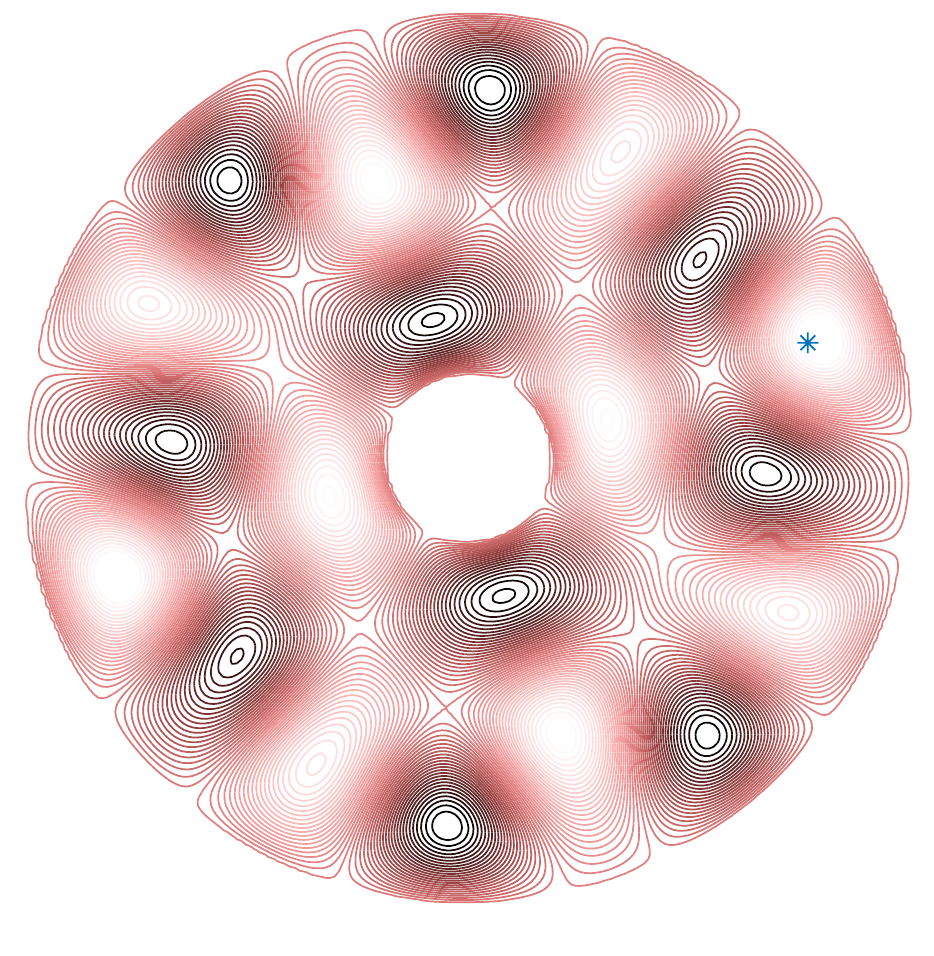}\hspace{.7cm}
\includegraphics[trim = 6cm 10cm 5.5cm 10cm, clip, scale=0.55]{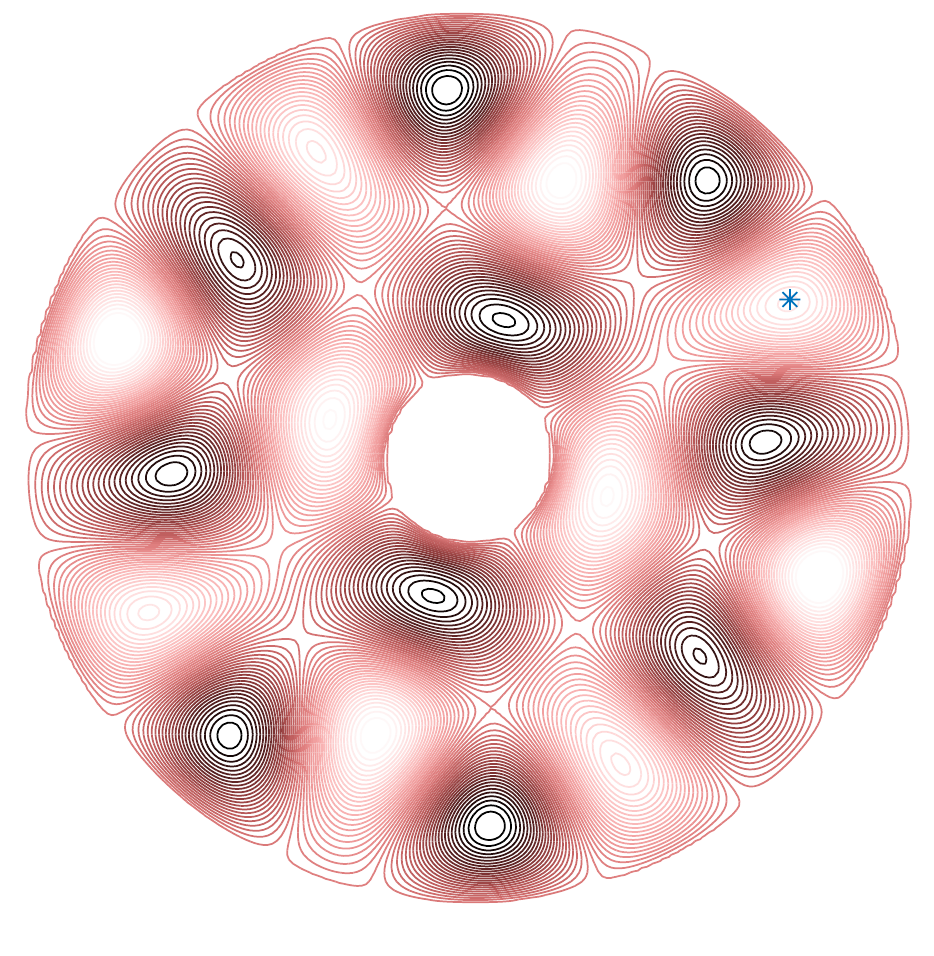}}
\centerline{\includegraphics[trim = 6cm 10cm 5.5cm 10cm, clip, scale=0.55]{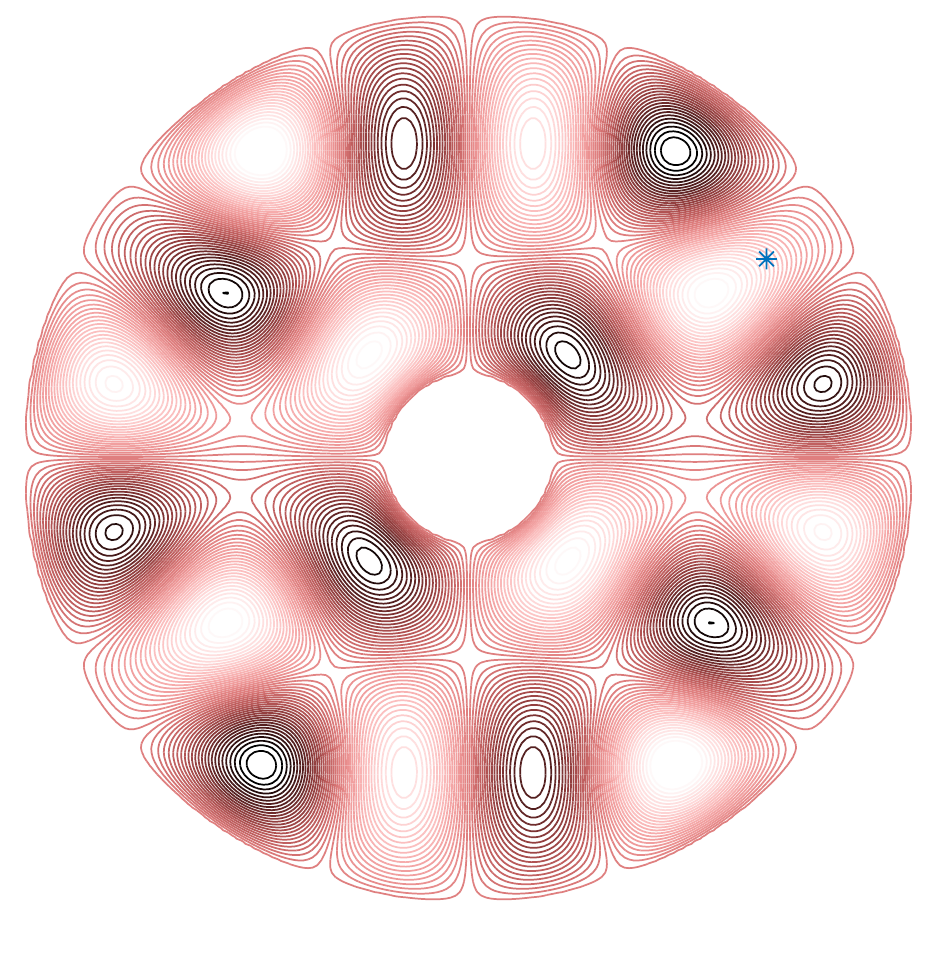}\hspace{.7cm}
\includegraphics[trim = 6cm 10cm 5.5cm 10cm, clip, scale=0.55]{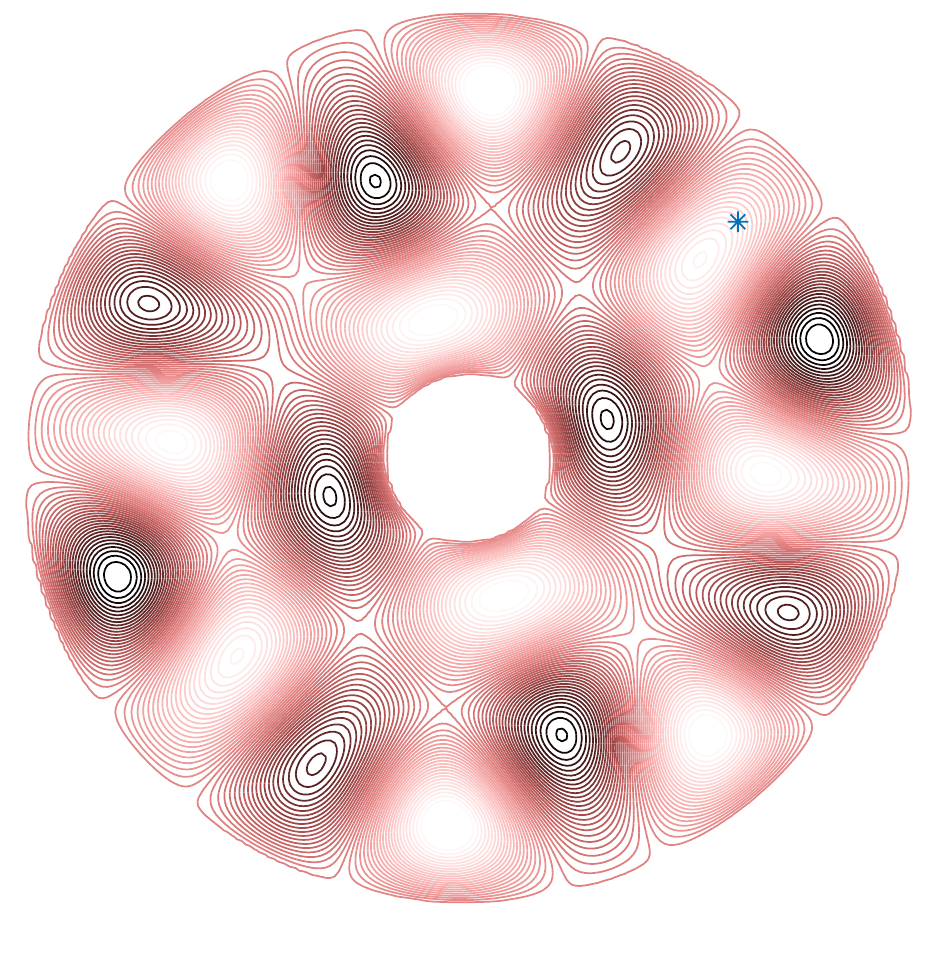}}
\vspace{.1cm}
\begin{caption}{\small Solution of the wave equation for a period of time. As the central core makes a half
clock-wise rotation, the peripheral part accomplishes an eighth of a cycle in anti clock-wise manner, as testified
by the position of the asterisk.}
\label{giri}
\end{caption}
\end{figure}
\end{center}

By suitably adjusting $t_0$, $\alpha_m$ and $\beta_m$, we get interesting evolution patterns.
This is the case for instance of the plots of Fig. \ref{giri}, where $m_A=2$, $m_B=8$, $\omega \approx 2.39$ and $ r_{\rm max}/r_{\rm min}=5.35 $.
The combination of $\Phi_2$ and $\Psi_2$ gives origin to the part of the solution that rotates with azimuth $\zeta=c\omega t -2\phi$
and internal velocity $c\omega r_{\rm min}/2\approx 1.2c$. Similarly, the part related to $\Phi_8$ and $\Psi_8$  counter rotates with azimuth 
$\zeta=c\omega t +8\phi$
and external velocity $c\omega r_{\rm max}/8\approx  1.6c$. Note that if the body was rigid, the external velocity would have been 
approximately equal to $6.4c$, exaggeratedly exceeding the speed of light. Another example of this type is shown in Fig. \ref{cuscinetto},
where the parameters are $m_A=4$, $m_B=20$, $\omega \approx 13$ and $ r_{\rm max}/r_{\rm min}\approx 2 $.
Here the external velocity is even lower than the internal one.
These plots can be however fully understood and appreciated only with the help of  animations.

\vspace{.1cm}



\section{Validity in a more extended context}

By a suitable merging of the classical electrodynamics equations with those
ruling  inviscid fluids,  a revision of the theory of EM in vacuum has been proposed in \cite{funarol} 
and successively developed in \cite{funarol2}.
Such an alternative model strictly includes the subspace of solutions of the classical Maxwell's equations, so that it
allows for the study of a larger variety of electromagnetic phenomena.
The main extension is the possibility to deal with situations where the divergence
of the electric field is different from zero, even if actual charges are not present (as electrons or positive nuclei).
An important motivation in favor of the revised model relies on the possibility to build compact solutions (solitary waves) traveling undisturbed without dissipation (see (\ref{sol1})).


We can find many arguments to support the fact that the wave
equation for ${\bf E}$ and the divergence-free condition ${\rm div}{\bf E}=0$ cannot hold, in general,
at the same time. The most trivial one is that, in order to get  (\ref{waveeb}) we must use ${\rm div}{\bf E}=0$,
since $\ {\rm curl}({\rm curl}{\bf E})=-\Delta {\bf E}+\nabla({\rm div}{\bf E})$.
On the other hand, mathematically, any vector wave equation, furnished with initial and boundary conditions
(imposed in a quite arbitrary manner), has unique solution. The chances that such a solution has (in addition) zero
divergence, are rather limited. 
A fast movement of charged bodies may actually create regions in the intermediate vacuum, where
${\rm div}{\bf E}\not =0$. 


The revised set of equations reads as follows (check the similarity with those given for instance in \cite{jackson}, 
p. 491, concerning the modeling of plasmas):
\begin{equation}\label{sfem2}
\frac{\partial {\bf E}}{\partial t}~=~ c^2{\rm curl} {\bf B}~
-~\rho {\bf V}
\end{equation}
\begin{equation}\label{sfbm2}
\frac{\partial {\bf B}}{\partial t}~=~ -{\rm curl} {\bf E}
\end{equation}
\begin{equation}\label{sfdb2}
{\rm div}{\bf B} ~=~0
\end{equation}
\begin{equation}\label{slor2}
\rho\left( \frac{D{\bf V}}{Dt}~+~\mu ({\bf E}+{\bf V}\times {\bf
B})\right) ~=~-\nabla p
\end{equation}
where we defined $\rho ={\rm div}{\bf E}$, and:
\begin{equation}\label{sub}
\frac{D{\bf V}}{Dt}~=~\frac{\partial {\bf V}}{\partial t}~+~({\bf
V}\cdot \nabla ){\bf V}
\end{equation}
Relation (\ref{slor2}) is the 
Euler's equation for the velocity field ${\bf V}$, provided with a 
forcing term of
EM type given by the vector: ${\bf E}+{\bf V}\times
{\bf B}$. The Maxwell's case in vacuum is now recovered by imposing 
$\rho =0$
and $p=0$. Therefore, the new set of equations is actually an extension 
of the classical model. 
The case when $D{\bf V}/Dt =0$ and $p=0$ will be mentioned later in 
this section.


Dimensionally, the constant $\mu$ is charge divided by mass. In
Appendix H in \cite{funarol2} is estimated to be approximately equal to
$
2.85\times 10^{11}~{\rm Coulomb / Kg}
$, which is
more or less of the same order of magnitude of the elementary charge
divided by the electron mass.
Up to dimensional scaling, the scalar $p$ plays the role of pressure. 
Indeed, the quantity $(\epsilon_0 /\mu ) p$ (where $\epsilon_0$ is the dielectric 
constant in vacuum)
is dimensionally equivalent to a force per unity of surface.


A further equation, related to energy conservation arguments, is finally added:
\begin{equation}\label{presse}
 \frac{\partial p}{\partial t} ~=~\mu\rho ~{\bf E}\cdot {\bf V}
\end{equation}
This says that pressure may come into place as a consequence of a lack of
orthogonality between ${\bf E}$ and ${\bf V}$. This is definitely what happens
to the fields introduced in the previous sections.


By taking the divergence of  (\ref{sfem2}), it is straightforward to get the following continuity equation:
\begin{equation}\label{cont}
\frac{\partial \rho}{\partial t}+{\rm div}(\rho {\bf V})=0
\end{equation}
Thus, the above relation is a consequence of the model and does not represent an extra requirement.
Finally, by scalar multiplication of (\ref{slor2}) by ${\bf V}$ and by taking
into account~(\ref{presse}), we arrive at a Bernoulli's type equation:
\begin{equation}\label{bern}
\frac{\rho}{2}\frac{D \vert {\bf V}\vert^2}{Dt}~+~\frac{Dp}{Dt}~=~0
\end{equation}


Since the EM fields considered in the previous section satisfy 
the whole set of Maxwell's equations (in particular $\rho=\rho_D=0$), they are
also solutions of the extended model. We can get new solutions by adding suitable stationary (not depending on time)
fields ${\bf E}_S$ and ${\bf B}_S$. This can be easily done if we assume that $\rho_S={\rm div}{\bf E}_S$
is constant (in particular we may choose $\rho_S=q$, for a given $q$). Thus, we set:
${\bf E}={\bf E}_D+{\bf E}_S$, ${\bf B}={\bf B}_D+{\bf B}_S$,
$\rho=\rho_D+\rho_S=\rho_S=q$, $p=p_D+p_S$. In addition, we take ${\bf V}$ as in (\ref{vu}).
By plugging these expressions in (\ref{sfem2})-(\ref{slor2}), and recalling the relation (\ref{evb}),
we must have:
\begin{equation}\label{sfem2a}
c^2{\rm curl} {\bf B}_S =\rho_S {\bf V}
\end{equation}
\begin{equation}\label{sfbm2a}
{\rm curl} {\bf E}_S=(0,0,0)
\end{equation}
\begin{equation}\label{sfdb2a}
{\rm div}{\bf B}_S =0
\end{equation}
\begin{equation}\label{slor2a}
\rho_S\left( \frac{D{\bf V}}{Dt}~+~\mu ({\bf E}_S+{\bf V}\times {\bf
B}_S)\right) ~=~-\nabla p_S
\end{equation}
We can then  come out with an explicit expression for ${\bf E}_S$ and ${\bf B}_S$.
We first note that:
\begin{equation}\label{dvdt}
\frac{D{\bf V}}{Dt}=-\frac{c^2\omega^2}{m^2} \Big(r \sin^2\theta , \ r\sin\theta\cos\theta, \ 0 \Big)=
-\frac{c^2\omega^2}{2m^2} \nabla (r^2\sin^2\theta )
\end{equation}
Successively, a possible choice for the stationary fields is:
\begin{equation}\label{vs}
{\bf E}_S=\frac{q}{3}( r, \ 0, \ 0)=\frac{q}{6} \nabla r^2 \qquad \qquad\
{\bf B}_S=\frac{q\omega}{5mc}\Big( -r^2\cos\theta, \ 2r^2 \sin\theta, \ 0\Big)
\end{equation}
so that we get:
\begin{equation}\label{vbs}
{\bf V}\times {\bf B}_S=-\frac{q\omega^2}{5m^2}\Big( 2r^3\sin^2\theta , \ r^3 \sin\theta  \cos\theta, \ 0\Big)=
-\frac{q\omega^2}{10m^2} \nabla (r^4\sin^2\theta )
\end{equation}
Let us observe that the magnetic field written above is exactly the one generated by a rotating
sphere, uniformly charged (see e.g. \cite{griffiths}, example 5.11).
Putting all together we find out (see also (\ref{evb})):
\begin{equation}\label{pr}
p=- \frac{\mu q}{m\omega}( rH^\prime +H)S_2 \ \sin\theta \cos\zeta +
\frac{c^2\omega^2 q}{2m^2}r^2\sin^2\theta  -\frac{\mu q^2}{6} 
r^2 +\frac{\mu\omega^2 q^2}{10m^2 } r^4\sin^2\theta 
\end{equation}
Other (singular) stationary fields compatible with the set of equations are the following ones
(up to multiplicative constants):
\begin{equation}\label{cvs}
{\bf E}_S=\Big( \frac{1}{r^2}, \ 0, \ 0\Big) \qquad\qquad
{\bf B}_S=\Big( \frac{2\cos\theta}{r^3}, \  \frac{\sin\theta}{r^3}, \ 0 \Big)
\end{equation}
In this case we get:  ${\rm div}{\bf  B}=0$,  ${\rm curl}{\bf  B}=0$,
${\bf  V}\times{\bf  B}=(c\omega /m)\nabla (r^{-1}\sin^2\theta )$ and $\rho_S=0$.
These last fields can be trivially incorporated in the pure Maxwellian part.

Some connections with the standard MHD can also  be established. For example,
our solution satisfies the {\sl induction equation}:
$$
\frac{\partial {\bf B}}{\partial t}=\frac{\partial}{\partial t}({\bf B}_D+{\bf B}_S)=\frac{\partial {\bf B}_D}{\partial t}
=-{\rm curl}{\bf E}_D
$$
\begin{equation}\label{indu}
={\rm curl}({\bf V}\times {\bf B}_D) ={\rm curl}[{\bf V}\times ({\bf B}_D+{\bf B}_S)]={\rm curl}({\bf V}\times {\bf B})
\end{equation}
where, in the order, we used that ${\bf B}_S$ is stationary, and then (\ref{sfbm2}), (\ref{evb}), (\ref{vbs}).
The literature on exact solutions in MHD is quite rich. We just mention a few references: \cite{golovin}, \cite{picard}, \cite{shokri}.
The approach used in this paper is however rather original.


We conclude with some final remarks.
A special subset of solutions (called {\sl free-waves}) is obtained by simplifying (\ref{slor2})
in the following way:
\begin{equation}\label{slor1}
{\bf E}+{\bf V}\times {\bf B} ~=~0
\end{equation}
Thus, the Euler's equation will be trivially satisfied by setting: $D{\bf V}/Dt =0$ and $p=0$.
In this case it is simple to deduce the orthogonality relations: ${\bf
E}\cdot {\bf B}=0$ and ${\bf E}\cdot {\bf V}=0$. 
Moreover, we assume that the intensity of ${\bf V}$
is  constantly equal to the speed of light, i.e.: $\vert {\bf V}\vert =c$.
If ${\bf V}$ turns out to be the gradient of a
potential function $\Psi$, we also get: $\vert \nabla\Psi\vert =c$,
which is the stationary {\sl eikonal} equation, ruling the dynamics of wave-fronts in geometrical optics.
Hence, the revised set of equations strongly connects electrodynamics with classical optics,
an important achievement that goes far beyond the usual Maxwell's model.
It is also  easy to check that free-waves satisfy the induction equation for an ideal conducting fluid, 
establishing another important link with MHD.
In order to prove this, it is enough to take the curl of  (\ref{slor1}) and use (\ref{fbm1}).


Numerous examples of free-waves are available. Here, we just mention the following one in Cartesian coordinates
$(x,y,z)$:
\begin{equation}\label{sol1}
{\bf E}=\Big(0,~0,~cf(z)g(ct-x)\Big),\ \ \
{\bf B}=\Big(0,~-f(z)g(ct-x), ~0\Big),\ \ \ {\bf V}=(c,~0,~0)
\end{equation}
where $f$ and $g$ can be (almost) arbitrary. This wave solves (\ref{sfem2}), (\ref{sfbm2}), (\ref{sfdb2}), (\ref{slor1}) and shifts at the
speed of light along the $x$-axis. In general we have $\rho \not =0$.  
Imposing $\rho=0$ implies that $f$ is a constant function, showing that parallel
wave-fronts of Maxwell's type (${\rm div}{\bf E}=0$) can only be plane-waves of infinite extension.
Instead, if $f$ and $g$ have a prescribed bounded support and vanish at its boundary, the wave remains
constrained in a shifting portion of space. As we already said, the continuity equation is automatically satisfied.

\section{Speculations about the constitution of the Sun and the solar system}

Since the pioneering papers of H. Alfv\'en (see e.g. \cite{alfvenL}), the study of the evolving plasma in the heliosphere
 is a widely investigated subject. 
Moreover,  the role of plasma is recognized to be a primary factor to understand our universe
at all  scales of magnitude (see e.g. \cite{peratt2}, \cite{peratt}).
The EM environment introduced in the previous sections may represent a possible background distribution,
in support of more complex phenomena. If such a  guess is correct, we can come out with some
interesting observations.

Actually, we begin with advancing some conjectures about our Sun. We refer to the circular electric patterns
of  Fig. \ref{sfera}.
Assuming that the {\sl solar cells} have an averaged diameter $d= 1100$ Km, there are 
about $2\pi R_\odot/d \approx 4000$
of them along the equatorial circumference ($R_\odot $ being the Sun radius). This means that $m\approx 2000$.
We can choose $\ell$ in such a way that $\ell /m\approx 2$. With this choice, the number of cells lined up along the equator is
approximately equal to those lined up along a meridian. Roughly, the Bessel's function $J_\alpha (r)$ has its
first positive root for $r \approx \alpha$. The function is practically zero, presenting a sudden bump just
before such a root. This says that the cells
have a relatively small depth. In the case of $J_{\ell +1/2}(\omega r)$ (see (\ref{besseljyg})), we
then get $\omega\approx \ell /R_\odot$. According to (\ref{vu}), the intensity of ${\bf V}$ on the equator is 
$\vert {\bf V}\vert =c\omega R_\odot/m \approx c \ell /m \approx 2c$. Therefore, this final result is almost independent
of all the parameters, with the exception of $c$. We recall that $c$ appears in the wave equations  (\ref{waveeb}).

The solar sphere is a medium containing material particles, whose movement  is accompanied by their EM interactions.
Particles supply the EM field in their motion and, at the same time, they are dragged by a mechanism
related to Lorentz's force. Due to the fact that they are massive, the velocity constant $c$ should be suitably reduced, 
by arguing that the medium intrinsically presents a relative dielectric constant higher than that of vacuum, forcing the 
information described by ${\bf V}$ to evolve at lower velocities. A quantitative analysis (too technical for the purpose of this paper)
involves the knowledge of the electrical conductivity $\sigma$
of the Sun (see e.g. \cite{stix}, section 8.1.2).
If $c$ is the speed of light in vacuum, a period of rotation around the vertical axis, turns out to be approximately 
4.65 seconds. This is $16\times 10^3$ times smaller than the revolution period of the Sun of about 27 days.
Thus $c$ must be reduced accordingly.  

We can provide an alternative explanation. Instead of adapting the value of $c$ to the conductive characteristics of the solar plasma, 
we can continue to suppose that $c$ is the speed of light in vacuum. Therefore, there is a high-frequency pure EM wave turning around that acts
as a forcing term.
Such a wave may be rather simple as in Fig. \ref{sfere}. As charged particles are present, they are dragged into a rotatory motion,
but they do this by following patterns that are strictly related to various physical quantities, such as: the intensity of the charges involved, their masses, 
their density within the plasma. The slower global motion is a consequence of the above restrictions, whereas the EM information still develops
at its classical speed. This viewpoint stimulates a further conjecture. A star is formed when, due to the creation of a swirl in the EM background 
(like a tornado in air), preexisting particles glue together (by electrodynamical and gravitational forces) conferring stability to the newborn
 structure and finding  a state of dynamical equilibrium. We also observe that vortexes of electric type on
 the solar surface may give raise to magnetic loops ({\sl spicules}), as a trivial consequence of Faraday's law.
These filaments, that can carry particles as well, ignite the mechanism at the origin of solar flares. 
Of course, our construction is elementary if compared to the complexity of a star. On the other hand, we
are just building our assumptions based on the solutions of linear problems.

Outside the massive bulk of the star
we still have plasma, but with an extremely small concentration of particles. We are basically in vacuum
so that the information now really evolves at  speeds comparable to that of light. By this we do not mean that particles 
necessarily travel fast. It is the flow of EM information in which they are embedded in that develops at luminous velocities.

The Sun has several ways to let us know its presence. First, it emits photons. These tiny energy packets escape as a consequence of chemical or subatomic
reactions. They carry away EM energy and they are fully described by the model equations of section 6 (see
the fields expressed by  (\ref{sol1})), and, incidentally, also solve the MHD equations. Photons
 constitute the visible part, since they can be detected with our eyes or instruments. 
We claim that there is another mean, only indirectly observable, used by the Sun to leave fingerprints on the
surrounding space. The turbulent EM status of the star induces the creation of
complicated (but well organized) whirls and spirals as described in section 5.
 This process generates a sequence of encapsulated shells, whose size  reasonably grows geometrically.
  Inside each shell there are trapped 
EM waves,  coordinately traveling and performing a peculiar dance. The transition between a shell and the next one
happens with continuity.  Differently from \cite{colburn}, we have shown here that it is possible to connect the different domains by 
avoiding shocks on the magnetic fields. We also assumed
that the interfaces are surfaces displaying zero magnetic field, though this hypothesis
may be reviewed at the occurrence. These systems evolve at an averaged speed comparable to that of light.
As they become larger, the angular velocity diminishes. 
We also observe that rotating EM solutions constrained in finite regions of space
are well suited for domains having annular topology  (\cite{chinosi}, \cite{funaro4},
\cite{funaro6}). This may suggest developments based on other geometries.
In reality, changes in the magnetic field 
have been detected by interplanetary probes. Some theories have been consequently developed (see
\cite{siscoe}, \cite{burlaga},  \cite{burlaga2}). Quick intermittent magnetic reversals in proximity of the Sun 
have been also reported (see for instance: \cite{horbury}, \cite{bale}, \cite{kasper}). 


Thus, according to our viewpoint, there is an organized set of shells which is not directly
visible. We can however appreciate its existence in indirect way. Perhaps, this construction may contribute to explain the formation of
planets, initially in a state of fragments ({\sl planetesimals}) and successively compacted by self-gravity and  the action 
of an organized plasma (see e.g. \cite{alfvenL}, \cite{alfven2}). The EM storms surrounding the Sun, force the selection of distant regions
of space where bunches of wrecks may meet and join together.
Indeed, the analysis of the first modes involved in the construction of the external shells (consider the case $m=\ell$
examined in section 5) says that a privileged direction is that of the equatorial plane. This suggests a possible
explanation of the (almost) coplanar distribution of the planets. Moreover, it seems that there are more chances to find matter
in zones where the interface magnetic field vanishes. Indeed, according to \cite{karsten}, charged particles
actually tend to accumulate in regions where the magnetic field is of weakest strength.
Due to the geometrical growth of the shells, we can advocate for the existence of specific
spots where it is more likely to find planets. This turns out to be in agreement with the Titius-Bode law,
in which the averaged distance of the planets from the Sun follows a geometric growth rate:  $ .4 + .3\times 2^k \ $,  
where the unity of measure of the distance is expressed in AU.
Note that in the right picture of Fig. \ref{cuscinetto}, the ratio $r_{\rm max}/r_{\rm min}$ is actually very close to 2. 

Finally, we would like to emphasize another aspect of the construction  here proposed. The solutions may actually
contain a dynamical part plus a stationary one (or pseudo-stationary, by meaning that this develops at
speeds that are much lower than that of light). We illustrated how this can be done in section 6 with the
fields, ${\bf E}={\bf E}_D+{\bf E}_S$, ${\bf B}={\bf B}_D+{\bf B}_S$, although we do not have more
fancy analytic expressions  to show. Thus, the EM activity at the exterior of the Sun
can be enriched by the addition of pseudo-stationary components. In this way we simulate a kind of 
{\sl solar wind},  developing with continuity across 
the various layers. The line of force corotate with the Sun and may be either closed or open loops depending on the distance  
(see \cite{ferraro}, \cite{pneuman}). To this extent, we characterize the
idea of a {\sl Parker's spiral} (see \cite{parker}), as  a further message imprinted on the plasma.
Exact solutions of this type, in the context of MHD, are given for instance 
 in \cite{asghar}.  A laboratory recreation of a spiraling wind,
through a rapidly rotating plasma magnetosphere, has been achieved  in \cite{peterson}. 
Analytic solutions for the spiraling fields generated by a rotating magnetic dipole are given in
\cite{sari}.

Similar constructions can be applied to satellites. Indeed, each planet is expected to be surrounded by
 an EM environment similar to that studied so far.
Such an alteration comes into conflict with that of the Sun, producing a deviation of the solar
wind and the reconnection of the magnetic streamlines along alternative patters (see e.g. \cite{priest}). 
Changes in the Earth magnetic field are documented since long time (see e.g. \cite{fleming}). They occur
with variable periods (daily, seasonal, secular). The consequence of our reasoning is that our planet, during the
travel along its orbit, is at the mercy of complex preexisting periodic EM recurrences filling up the whole heliosphere
(see \cite{mcp}, \cite{milan}). We stop however our
exposition at this point. A deeper analysis on these issues would bring us too far from the limits we fixed for this paper.

\end{document}